# IMPROVED LIMITS ON GAMMA RAY BURST REPETITION [1]

Max Tegmark[1,2], Dieter H. Hartmann[3], Michael S. Briggs[4], Jon Hakkila[5]
& Charles A. Meegan[6]

[1] *Max-Planck-Institut für Physik, Föhringer Ring 6, D-80805 München;*
*max@mppmu.mpg.de*

[2] *Max-Planck-Institut für Astrophysik, Karl-Schwarzschild-Str. 1, D-85740 Garching*

[3] *Dept. of Physics and Astronomy, Clemson University, Clemson, SC 29634;*
*hartmann@grb.phys.clemson.edu*

[4] *Dept. of Physics, University of Alabama, Huntsville, AL 35899*

[5] *Dept. of Physics & Astronomy, Mankato State University, Mankato, MN 56002*

[6] *NASA/Marshall Space Flight Center, Huntsville, AL 35812*

### Abstract

We tighten previous upper limits on gamma ray burst repetition by analyzing the angular power spectrum of the BATSE 3B catalog of 1122 bursts. At 95% confidence, we find that no more than 2% of all observed bursts can be labeled as repeaters, even if no sources are observed to repeat more than once. If a fraction $f$ of all observed bursts can be labeled as repeaters that are observed to burst $\nu$ times each, then all models with $(\nu - 1)f \geq 0.05$ are ruled out at 99% confidence, as compared to the best previous 99% limit $(\nu - 1)f \geq 0.27$. At 95% confidence, our new limit is $(\nu - 1)f \geq 0.02$. Thus even a cluster of 6 events from a single source would have caused excess power above that present in the 3B catalog. We conclude that the current BATSE data are consistent with no repetition of classical gamma ray bursts, and that any repeater model is severely constrained by the near perfect isotropy of their angular distribution.

---

[1] Submitted to the Astrophysical Journal Letters

# 1 INTRODUCTION

The origin of cosmic $\gamma$-ray bursts (GRBs) is not known, and their distance scale has not been established. The angular isotropy of GRBs provides an important clue which has generated a "great debate" about the question whether GRBs are of galactic or cosmological origin (e.g., Briggs 1995; Fishman 1995; Fishman & Meegan 1995; Hartmann 1995a; Lamb 1995; Paczyński 1995). While cosmological models usually invoke singular events, such as the merger of two compact objects (e.g., Narayan, Paczyński & Piran 1992; Meszaros & Rees 1993), Galactic models currently under consideration require multiple outbursts from each source. Recurrence in the framework of cosmological models could occur due to lensing, but the frequency of such events should be very small (e.g., Nemiroff *et al.* 1994). Detection of a significant fraction of repeating GRBs would argue against a cosmological origin and favor a Galactic (halo) origin. It would definitely exclude cosmological models in which the source is destroyed.

To satisfy the isotropy constraint, a Galactic halo must be very large in order to minimize the dipole due to the solar offset from the Galactic center. The current multipole limits (Briggs *et al.* 1995; Tegmark *et al.* 1995) require galactocentric shells with typical radii $\sim 200$ kpc. On the other hand, halos that are too large will yield an excess of bursts towards M31, which is not observed (e.g., Hakkila *et al.* 1994, 1995; Briggs, *et al.* 1995). Because of these twin constraints, most halo models invoke a limiting sampling distance of about 300 kpc for the BATSE bursts.

Currently, the only surviving Galactic model invokes bursts that are produced by high velocity pulsars (HVPs) born in the vicinity of the disk streaming out into the halo with velocities $\sim 10^3 \,\mathrm{km\,s^{-1}}$ (Li & Dermer 1992; Duncan & Thompson 1992; Duncan, Li & Thompson 1993; Woosley 1993; Li, Duncan & Thompson 1994; Colgate & Leonard 1994, 1995; Li & Duncan 1995; Bulik & Lamb 1995; Podsiadlowski *et al.* 1995; Woosley & Herant 1995). The recent upward revision of radio pulsar velocities (Lyne & Lorimer 1994) provides some support and motivation for such a scenario.

Radio pulsars are born in the Galaxy at the rate of roughly one pulsar every 100 years (e.g., Narayan & Ostriker 1990). This is consistent with recent estimates of the Galactic supernova rate ($2.5\pm0.5\ 10^{-2}\ \mathrm{yr}^{-1}$; Tammann, Löffler & Schröder 1994) if a significant fraction of them leave a black hole instead of a neutron star. A typical value for the fraction of pulsars with velocities $\gtrsim 10^3 \,\mathrm{km\,s^{-1}}$ is $\sim 10\%$ (Lyne & Lorimer 1994; Frail, Goss & Whiteoak 1994). However, not every HVP may become a GRB source if additional selection criteria, such as particular magnetic field strengths, must be applied. The fraction of pulsars with sufficient energy, if rotational momentum is the energy source, is probably much smaller than unity (Hartmann & Narayan 1995). On the other hand, BATSE has been detecting GRBs at the rate of approximately one per day. After correcting for Earth blockage and temporal gaps, the inferred all-sky GRB rate at the BATSE sensitivity level is $\sim 10^3$ yr$^{-1}$. The comparison of the pulsar birth rate with the high observed rate of bursts implies that burst sources must repeat in the HVP scenario (Hartmann & Narayan 1995; Lamb 1995; Podsiadlowski, Rees & Ruderman 1995). While the actual detection of burst recurrence is perhaps still avoidable, the parameter range of realistic halo models clearly encourages the search for repeaters in the data. On the other hand, the leading cosmological scenario of merging compact objects would be in serious trouble if even a small number of GRBs could convincingly be shown to originate from the same source.

Do classical GRBs repeat, and if yes can we determine their recurrence pattern? The search for recurrence of classical GRBs (in this work we exclude the exciting class of soft gamma repeaters, which were recently reviewed by Kouveliotou (1994)) has a long tradition (Mazets, *et al.* 1981; Schaefer & Cline 1985; Atteia *et al.* 1985, 1987). Though somewhat model dependent, the distribution of burst locations studied by these authors suggests a lower limit to the GRB recurrence time of $\sim 10$ years. Interest in this topic was revived by Quashnock & Lamb (1993), who found evidence for repetition in the BATSE 1B dataset (Fishman *et al.* 1994) using the the nearest neighbor (NN) statistic (e.g., Scott & Tout 1989).

Recurrence, even in a single case, would be immediately obvious if we had perfect locations. The locations provided by BATSE, while numerous, are imperfect and consequently a statistical analysis is required to demonstrate or limit the presence of repeaters. The positions in the first BATSE catalog had a minimum uncertainty of $\sim 4°$ due to systematics (Fishman *et al.* 1994). The repetition analysis of Quashnock & Lamb has been controversial. For example, Narayan & Piran (1993, 1994) used an apparent excess of burst pairs with $\sim 180°$ angular separation ("antipodal bursts") to argue against the repeater hypothesis derived from the small-angle NN excess, but Quashnock & Lamb (1994) argued that real physical (Galactic) anisotropies in the catalog are responsible for the positive antipodal correlations.

In contrast to BATSE, COMPTEL and EGRET can localize bursts to $\sim 1°$, but the event rate of these detectors is very small. However, the recent near coincidence of COMPTEL bursts GRB930704 and GRB940301 (Kippen *et al.* 1995a,b) suggests repetition, because in three years of operation such a coincidence had only a 3% chance probability. The most recent compilation of 27 GRB positions observed with COMPTEL (Kippen *et al.* 1995c) did not yield another pair of coincident bursts, implying a reduced significance of the first pair.

While most studies focused on projected GRB positions, it is clear that there could also be a clustering effect in the time domain. This aspect was first investigated by Wang & Lingenfelter (1994, 1995ab), who also found evidence for repetitions.

Given the importance of these findings for burst models, confirmation in subsequent samples is essential. Angular correlation function studies as well as nearest neighbor methods applied to the 2B sample did not confirm the earlier claims and instead found the data to be consistent with no repetition (Blumenthal *et al.* 1994; Meegan *et al.* 1995a) The small scale excess was reduced and the antipodal excess also went away (Hartmann *et al.* 1994). However, the 2B data suffered from the problem of large data gaps in time due to the failure of the tape recorders aboard CGRO. The resulting lower exposure to bursts obviously reduces our ability to detect recurrent events. Taking these effects into account, Meegan *et al.* (1995a) constrain the total fraction of repeaters among the 585 observed 2B bursts to $f \leq 20\%$. This limit is based on simple repeater pairs in all cases. If each repeating source produces more than 2 detectable events, the corresponding limit would be tighter. The data also do not show evidence for enhanced clustering in time/space (Brainerd *et al.* 1995), although different statistical measures, such as the Mantel-Haenszel test, provide marginal evidence for joint temporal and angular clustering with a time scale of 4–5 days (Petrosian & Efron 1995).

A comparison of the nearest neighbor method and the angular correlation function method shows that stronger constraints are obtained from the latter (Meegan *et al* 1995a; Brainerd 1995). These findings contradict earlier statements that the NN statistic is the superior method for studies of small scale clustering (Lamb, Quashnock, & Graziani 1993). Many authors have introduced new statistical techniques— most have concluded that there is either weak or no evidence for burst repetition (Hartmann *et al.* 1995; Bennett & Rhie 1995a; Efron & Petrosian 1995; Hurley *et al.* 1994).

To make further progress in the analysis of angular distribution data it is necessary to remove the two most important obstacles in this analysis; data gaps and poor positions. The most recent set of BATSE data (the 3B catalog; Meegan *et al.* 1995bc) is free of data gaps other than those induced by the South Atlantic Anomaly passages, and a major effort to improve the positioning algorithm (Pendleton *et al.* 1995) has reduced the systematic error to 1.6°. This sample of bursts thus provides a solid base for tests of the repeater hypothesis. Analyses of the 3B data using the standard tools of NN statistic and angular correlation functions (Meegan *et al.* 1995c) confirmed the conclusions derived from the 2B data: classical GRBs have not been observed to repeat. The same conclusion was reached in a study of 3B data using the matched pair statistic (Bennett & Rhie 1995b). While supporting the null hypothesis, these studies did not significantly improve the limits on the repeater fraction. It is the purpose of this *Letter* to introduce a new statistical method that provides significantly more statistical power than the standard tools. We apply the method to the 3B catalog and derive improved limits on the repeater fraction.

The remainder of this *Letter* is organized as follows: in §2 we present our repeater statistic, and in §3 we apply it to the 3B data set and discuss the results.

## 2 METHOD

As in Meegan *et al.* (1995a), we find it convenient to work with a two-parameter family of repeater models. These models are specified by the parameters $f$ and $\nu$, where $f$ is the fraction of all observed bursts that can be labeled as repeaters and $\nu$ is the average number of observed events per source observed to repeat. Our comparison of these models with the data will proceed as follows:

1. We select a method of reducing an entire data set into a single number $R$, a number which is sensitive to the type of burst clustering that repetition would produce. We chose the sign so that the more evidence there is for repetition in the data set, the larger $R$ will be.

2. For fixed values of $f$ and $\nu$, we compute the probability distribution of $R$ by making Monte Carlo simulations of mock BATSE 3B catalogs.

3. We compute the observed value of $R$, denoted $R_{\mathrm{obs}}$, from the real BATSE 3B catalog.

4. Combining the results of the two previous steps, we obtain the function $p(f,\nu)$, defined as the probability that $R \leq R_{\mathrm{obs}}$ given that the true parameter values are $f$ and $\nu$. For instance, if $p(f,\nu) = 5\%$, then we would expect to observe as low an $R$-value as we did, by chance, only 5% of the time, and conclude that the model $(f,\nu)$ is ruled out with 95% confidence.

5. By repeating this analysis for a grid of points in parameter space and making a contour plot of $p(f,\nu)$, we obtain our final results, shown in Figure 2.

Clearly, the success of this approach depends crucially on the choice of the statistic $R$. Any choice whatsoever will of course give statistically valid results as long as the mock catalogs and the real data are treated in the same way, but poor choices of $R$ will not allow good discrimination between repeating and non-repeating models. The best choice of $R$ is clearly that which allows us to rule out as many incorrect models as possible, *i.e.*, in statistics jargon, that which gives our statistical test the maximum power. Many repeater statistics have already been employed in the literature, as discussed in §1. Currently, the standard tests use the nearest neighbor test or the two-point correlation function. These are both good choices, as they are sensitive to small-scale clustering of the type that repetition produces. Here we will use another choice, which we will argue is still better.

## 2.1 The total power statistic

In Tegmark, Hartmann, Briggs & Meegan (1995, hereafter THBM95), a method was presented for computing the angular power spectrum $C_\ell$ of gamma ray bursts in the presence of the position errors of BATSE. It was found that in terms of the power spectrum, burst repetition has a very simple signature: the power at all multipoles $\ell$ is increased by the same amount. Therefore a logical measure of burst repetition would be the sum (or, apart from an irrelevant multiplicative constant, the average) of the power in all multipoles, $R = \sum_{\ell=0}^{\infty} C_\ell$, i.e., the *total power*. However, the position errors make the estimates of high multipoles very noisy, and in THBM95 it was found that the shot noise error bars on the power spectrum estimates explode for $\ell \gtrsim 70$. To be useful, $R$ should be fairly insensitive to noise, so we clearly want to give less weight to the $C_\ell$ with large error bars, *i.e.*, with large $\ell$-values. With this in mind, we propose the following repeater statistic:

$$R \equiv \sum_{\ell=0}^{\infty} \sum_{m=-\ell}^{\ell} \left( \frac{N_\ell^{\text{eff}}}{N} \right)^2 |\tilde{a}_{\ell m}|^2. \qquad (1)$$

As was shown in THBM95, the minimum-variance estimate of the spherical harmonic coefficient $a_{\ell m}$ when faced with location uncertainties is

$$\tilde{a}_{\ell m} \equiv \frac{N}{N_\ell^{\text{eff}}} \int Y_{\ell m}(\hat{\mathbf{r}}) x(\hat{\mathbf{r}}) d\Omega. \tag{2}$$

Here $x$ is the *smoothed burst map*, plotted in THBM95, which is simply a sky map of all bursts smeared out by their position uncertainties:

$$x(\hat{\mathbf{r}}) = \sum_{k=1}^{N} B_k(\hat{\mathbf{r}}), \tag{3}$$

where in the approximation of a Gaussian beam function,

$$B_k(\hat{\mathbf{r}}) = \frac{\exp\left[-\frac{1}{2}\frac{\theta^2}{\sigma_k^2}\right]}{2\pi\sigma_k^2}, \tag{4}$$

where $\theta$ is the angle between $\hat{\mathbf{r}}$ and $\hat{\mathbf{r}}_k$, the position of burst $k$. Up to an irrelevant additive constant (the shot noise $b_{\ell m}$ discussed in THBM95), $|\tilde{a}_{\ell m}|^2$ is a measure of the power $C_\ell$. Just as in THBM95,

$$N_\ell^{\text{eff}} \equiv \sum_{k=1}^{N} e^{-\sigma_k^2 \ell(\ell+1)}, \tag{5}$$

where $\sigma_k$ is the uncertainty in the position of burst $k$, and gives the effective number of bursts that are well enough localized to contribute information about $C_\ell$. In THBM95, the error bars on $C_\ell$ were found to scale as $1/N_\ell^{\text{eff}}$, so the weights $(N_\ell^{\text{eff}}/N)^2$ in our definition of $R$ have the desired property of suppressing the influence of the noisy high-$\ell$ part of the power spectrum, since $N_\ell^{\text{eff}} \to 0$ as $\ell \to \infty$.

## 2.2 A faster way to compute the total power

Since the quantity $R$ is a measure of the total fluctuation power in the burst distribution, which is a rather natural quantity, one may ask if there is a simpler way of computing it which circumvents time-consuming calculations of large numbers of spherical harmonics $Y_{\ell m}$. Fortunately, the answer to this question turns out to be yes. Substituting equation (2) into equation (1) and using the spherical harmonic identity

$$\sum_{\ell=0}^{\infty} \sum_{m=-\ell}^{\ell} |x_{\ell m}|^2 = \int |x(\hat{\mathbf{r}})|^2 d\Omega \tag{6}$$

(the spherical version of Parseval's theorem), we obtain the simple and useful result

$$R = \int x(\hat{\mathbf{r}})^2 d\Omega. \tag{7}$$

Equation (7) provides an intuitive way of understanding how the statistic $R$ works. If all bursts are well separated, so that their respective Gaussians hardly overlap, then $R$ will simply be a sum of separate contributions from each burst, and will be independent of the exact burst positions. If clustering is present, however, then we get "constructive interference" where two Gaussians overlap, and since the integrand in equation (7) is squared, a larger contribution than if they did not overlap. Also, since the Gaussians $B_k$ are normalized as probability distributions (they integrate to unity), their peak values are larger for well-localized bursts. This means that overlaps contribute more to the sum if they involve a well-localized burst (with a small $\sigma_k$).

## 2.3 A still faster way

We found that, $4\pi$-factors aside, the R-statistic is simply the mean squared amplitude of the smoothed burst map. Since we wish to make many thousands of Monte-Carlo simulations to obtain accurate estimates of the probability distributions of step 2, it is desirable to further accelerate the procedure of computing $R$. Substituting equation (3) into equation (7), we obtain

$$R = \sum_{i=1}^{N} \sum_{j=1}^{N} \int B_i(\hat{\mathbf{r}}) B_j(\hat{\mathbf{r}}) d\Omega. \tag{8}$$

In the approximation that $\sigma_k \ll$ one radian $\approx 60°$ (which is quite an accurate approximation, as typical values are a few degrees), the integral reduces to that of the product of two Gaussians in the flat two-dimensional plane, and can be done analytically. This leaves us with the handy result

$$R = \frac{1}{2\pi} \sum_{i=1}^{N} \sum_{j=1}^{N} \frac{\exp\left[-\frac{1}{2} \frac{\theta_{ij}^2}{(\sigma_i^2 + \sigma_j^2)}\right]}{\sigma_i^2 + \sigma_j^2}, \tag{9}$$

where $\theta_{ij} \equiv \cos^{-1}(\hat{\mathbf{r}}_i \cdot \hat{\mathbf{r}}_j)$, *i.e.*, the angle between the two bursts. Finally, since replacing our statistic $R$ by a monotonic function $f(R)$ (for instance rescaling $R$ and subtracting off a constant) will in no way change its ability to discriminate between models, let us redefine $R$ to give numbers of a convenient magnitude. If there is no clustering and the total area $\sim \sum_i \sigma_i^2$ covered by all the error circles is so small that substantial overlap is unlikely, then the sum (9) will be dominated by the "diagonal" terms with $i = j$ and reduce to $R \approx (1/4\pi) \sum_i \sigma_i^{-2}$, an expression which is completely independent of the burst locations $\hat{\mathbf{r}}_i$. Dividing by this quantity and disposing of uninteresting additive and multiplicative constants, we thus redefine our $R$-statistic as

$$R \equiv \left(\sum_{i=1}^N \sigma_i^{-2}\right)^{-1} \sum_{i=1}^N \sum_{j=1}^{i-1} \frac{\exp\left[-\frac{1}{2}\frac{\theta_{ij}^2}{(\sigma_i^2+\sigma_j^2)}\right]}{\sigma_i^2 + \sigma_j^2}, \qquad (10)$$

This is the expression that we use in our calculations. Apart from an irrelevant factor of 4, it is merely the ratio between the off-diagonal ($i \neq j$) and diagonal ($i = j$) elements in the sum (9). It will clearly always be non-negative, and in the absence of clustering, it will approach zero if the location errors $\sigma_i$ do.

The position uncertainties $\Delta\theta$ quoted in the BATSE 3B catalog are defined as the radius of the one sigma circle, *i.e.*, of the circle that contains $\text{erf}[1/\sqrt{2}] \approx 68\%$ of the probability. Thus in the limit $\sigma_k \ll 1$, the conversion between $\Delta\theta$ and $\sigma$ is

$$\frac{\sigma}{\Delta\theta} = \left[-2\ln\left(1 - \text{erf}\left[\frac{1}{\sqrt{2}}\right]\right)\right]^{-1/2} \approx 0.66. \qquad (11)$$

Note that the values of $\Delta\theta$ quoted in the BATSE 3B catalog do not include the systematic error contribution of $1.6°$, which is to be added to the quoted values in quadrature.

In defining our statistic $R$ above, we have omitted a few elements that were used in THBM95, for instance the spatially varying exposure function $\bar{n}$ and the Fisher beam function. It should be emphasized that despite these omissions and the various approximations made (that $\sigma_k \ll 60°$, *etc.*), our statistical statements will be 100% exact. This is because, as mentioned above, we are free to define the statistic $R$ however we want, as long as we make no approximations when generating the Monte Carlo catalogs, and compute $R$ in exactly the same way from these and from the real data. The only real constraint is that if we make $R$ depart too much from the exact measure of the total power, it may no longer be as good a measure of clustering and its ability to reject incorrect models will be weakened.

## 2.4 The mock catalogs

According to the model, the $N = 1122$ bursts where caused by $(1-f)N$ non-repeating objects (rounded to the nearest integer) and $fN/\nu$ repeating objects that burst $\nu$ times each. We therefore generate the mock BATSE 3B catalogs as follows:

1. The $(1-f)N + fN/\nu$ objects are distributed randomly across the sky, in a completely uncorrelated fashion, but with the point density modulated by the exposure function $\bar{n}(\hat{\mathbf{r}})$ as described below.

2. Each non-repeating object is assigned one burst and each repeating object is assigned $\nu$ bursts.

3. The 1122 position errors $\sigma_k$ from the BATSE 3B data set are re-sorted in a random order, and one assigned to each burst.

4. Each burst is displaced from its true position by a random amount drawn from the probability distribution $B_k$ of equation (4). (In fact, we use the more accurate Fisher beam function described in THBM95, but it is virtually identical for $\sigma \ll 60°$.)

We truncate the number of bursts from the last repeating source so that each mock catalog contains exactly 1122 bursts.

## 2.5 The exposure function

The sky exposure of BATSE is not quite uniform (Fishman *et al.* 1994; Meegan *et al.* 1995b). The BATSE experiment does not exclude any area of the sky, but due to blocking by the Earth and detector gaps during passages of the so called South Atlantic Anomaly, some positions on the sky have a reduced probability for burst detection. We quantify this by the function $\bar{n}(\hat{\mathbf{r}})$, the *exposure function*, defined as the expected number of bursts per steradian. It is simply proportional to the total exposure time that each patch of sky has received. It is known a priori and is not obtained from the observed burst distribution.

Because of problems due to the loss of the spacecraft tape recorders, the absolute efficiency has not been determined since the release of the 1B data set. However, the shape of the exposure function $\bar{n}$ is essentially independent of time, and since the shape is all that matters for the present analysis, we employ the 1B estimate (Fishman *et al.* 1994). This function $\bar{n}$ depends on declination only, and is independent of right ascension. This means that in equatorial coordinates, the multipole coefficients $\bar{n}_{\ell m}$ vanish except when $m = 0$. The dominant deviation from uniformity is a quadrupole ($\bar{n}_{20}/\bar{n}_{00} \approx 8.8\%$) depletion of bursts near the equator due to the shadowing of the sky by the earth. The second largest anisotropy is a dipole moment ($\bar{n}_{10}/\bar{n}_{00} \approx 4.5\%$) towards the earth's north pole, due to the South Atlantic Anomaly, which requires disabling triggers. Compared to the shot noise, the higher multipoles ($\ell \geq 3$) are negligible ($a_{\ell 0}/a_{00} \lesssim 1\%$), but for completeness, they have nonetheless been included in our analysis.

We incorporate the effect of variable exposure into our mock surveys as follows:

1. We compute the maximum value of the function $\bar{n}(\hat{\mathbf{r}})$ and denote it $\bar{n}_{\max}$.

2. Whenever a burst has been generated at a position $\hat{\mathbf{r}}$, we accept it with a probability $P \equiv \bar{n}(\hat{\mathbf{r}})/\bar{n}_{\max}$ and reject it with a probability $1 - P$. In other words, we generate a uniformly distributed random number $u \in [0, 1]$, and if $u > \bar{n}(\hat{\mathbf{r}})/\bar{n}_{\max}$, then we generate a new burst position $\hat{\mathbf{r}}$ and a new random number $u$, repeating the procedure until we obtain $u \leq \bar{n}(\hat{\mathbf{r}})/\bar{n}_{\max}$.

The net result is that, when averaging over many mock catalogs, there are on average $\bar{n}(\hat{\mathbf{r}})d\Omega$ bursts in a solid angle $d\Omega$ around $\hat{\mathbf{r}}$.

## 3 RESULTS AND DISCUSSION

Figure 1 shows the cumulative probability distribution of the $R$ statistic for a range of models with $\nu = 2$. For each of these models, we generated $10^4$ Monte Carlo catalogs from which we computed the $R$ statistic, producing a single curve in the figure. These cumulative probability distributions $F(R_*)$ show the fraction of the $R$-values that are smaller than any one constant $R_*$, so $F(R_*)$ is simply the integral $\int_0^{R_*}$ of the probability distribution for $R$. For instance, the median $R$-value is the point $R_*$ where $F(R_*) = 0.5$.

The first thing to notice about Figure 1 is that $R$ behaves as expected: as the repeater fraction $f$ increases, the distribution of $R$-values shifts further to the right. Secondly, the value extracted from the real data, $R_{\text{obs}} \approx 0.3085$, lies far to the left in the figure, which means that the data contains no evidence whatsoever for repetition. $F(R_{\text{obs}}) \approx 5\%$ for the model $\nu = 2$, $f = 0.02$, which means that this model produces such a low $R$ value only 5% of the time, *i.e.*, that this model is ruled out at 95% confidence. Models with higher $f$-values are of course even less consistent with the data.

At first sight, it may appear disturbing that $R_{\rm obs}$ is lower than the typical values obtained for the null model $f = 0$ which has no repetition at all. Even if there are no repeaters, $R$ will only be as low as 0.3085 a mere 12% of the time. Since no form of clustering could explain this (rather, a contrived model with some form of "anti-clustering" would be needed), one is lead to ask how unlikely it is that this merely happened by chance. The answer is of course that if the no repetition hypothesis is true, then the probability of finding an $R$-value this far out in one of the tails of the distribution is $2 \times 12\%$, *i.e.*, it would happen about a quarter of the time and thus should not be a source of concern. A similar conclusion was reached by Hartmann & Epstein (1989), who also found an unusually isotropic burst sample in their analysis of IPN positions from Atteia *et al.* (1987). In contrast, since we know a priori that repetition causes high rather than low $R$-values, we should use a single-sided (rather than double-sided) test when ruling out repeater models.

The results of repeating the Monte Carlo procedure for a two-dimensional grid of points in the $(f,\nu)$ parameter space are summarized in Figure 2. In agreement with Meegan *et al.* (1995a), the contours are seen to be accurately fit by the hyperbolic profile $(\nu - 1) \propto 1/f$. At 99% confidence, we have

$$(\nu - 1)f < 0.049, \tag{12}$$

and at 95% confidence, we obtain $(\nu - 1)f < 0.018$. For comparison, the contour of models ruled out at 99% confidence with the correlation function method applied to the same dataset is shown as a heavy solid line, $(\nu - 1)f = 0.27$ (Meegan *et al.* 1995c). It is seen that the method presented here significantly tightens the constraints on any observable repeater population. While no statistical analysis can disprove the possibility of a very low repetition rate, the constraints on simple repeater models discussed here are now so severe that either recurrence is very rare amongst classical GRBs (in contrast to the what earlier studies suggested), or it has a very special pattern we can no longer find in the 3B data although we did in 1B. The latter interpretation appears to be very contrived as long as we do not find theoretical support for such a special pattern.

The straight lines in Figure 2 correspond to models with a fixed number (1, 2, 4, 8 and 16) of repeating sources in the data set. For instance, the leftmost line corresponds to the case where all the repeated bursts are due to a single source, so that the repetition would manifest itself as a single cluster of $\nu$ bursts in the BATSE catalog. The fact that this line intersects the 99% contour below $\nu = 9$ means that our constraints are now so strong that not even a single repeater with $\nu = 9$ is allowed. Similarly, even a single sixfold repeater is excluded at 95% confidence.

## 3.1 How robust are the results?

How robust are these results to changing various assumptions about the data? Playing the devil's advocate, are there any types of errors that could produce artificially strong constraints? Most types of data problems, for instance neglected modulations in the exposure function, would have the opposite effect and give *weaker* upper limits on repetition, since they would create extra clustering in the data. There are basically only two exceptions:

1. If the true exposure function $\bar{n}$ is more smooth than assumed
2. If the location errors are underestimated

To quantify the effects of the first possibility, we re-ran the analysis with $\bar{n}$ constant. This is of course the most extreme case possible, since there is nothing more uniform than a uniform distribution, and quite contrived since we know that the South Atlantic Anomaly and Earth shadowing must have some modulating effects. Nonetheless, the changes were quite marginal, with the weakened limits being $(\nu - 1)f < 0.064$ at 99% confidence. In other words, a problem of type 1 could at most weaken the limits from 0.049 to 0.064, *i.e.*, by about 30%.

A more realistic source of concern is the second possibility. Graziani & Lamb (1995) analyzed the distribution of 3B locations in comparison to the positions derived from the IPN[3] network, and conclude that the systematic error $\Delta\theta_0$ of the 3B data should be $\sim 4°$ instead of the advertised 1.6°. In addition, there may be correlations in the data that suggest a brightness dependence for the systematic error, instead of the constant value suggested by Meegan *et al.* (1995c). The studies by Graziani & Lamb (1995) did not take systematic effects in the IPN localization method into account, and also do not incorporate the fact that some IPN locations are based upon the earlier BATSE 2B locations and thus may be biased against the 3B locations. Although we tend to agree more with the error budget prescribed by the BATSE Team, this is an unsettled question and we have therefore studied the effect of larger errors on our repeater limits. For simplicity, just to estimate the magnitude of the effects, we maintained a constant systematic error $\Delta\theta_0$ (as suggested for the 3B data) and repeated the entire analysis for $\Delta\theta_0 = 2°, 3°, 4°$ and $5°$. To be maximally conservative, we took $\bar{n}$ constant as well, and found that for $\nu = 2$, the 99% upper limit on the repeater fraction scaled approximately as

$$f < 0.06 \left(\frac{\Delta\theta_0}{1.6}\right)^{0.4}.$$

In other words, even if the the true systematic location errors were as high as $5°$, our limits would only be weakened by less than a factor of two.

## 3.2 Conclusions

In summary, we have sharpened previous limits on GRB repetition by analyzing the improved BATSE positions of the 3B catalog (Meegan *et al.* 1995bc) with a new statistic based on the angular power spectrum. This method is more powerful than the NN statistic or correlation functions because it is in a sense a global method (e.g., Hartmann *et al.* 1995). The presence of clustering (on any scale) somewhere on the sky reduces the density of sources everywhere on the sky (relative to an isotropic distribution). Any method that utilizes the impact of clustering on all angular scales is more sensitive to clustering than local methods, such as NN statistics that only measure neighbor excess very close to a given source, or angular correlations. The appearance of clusters other than those occurring by chance quickly leaves a mark on the overall, global angular power spectrum. Using this effect we find an amazing isotropy of bursts which is hard to satisfy with any model other than that of non repeating sources, providing strong evidence against the Galactic models currently under consideration. It was believed that spherical harmonic expansion would not provide a good tool for repeater studies (Lamb, Quashnock, & Graziani 1993), because the power is spread over many high harmonics. We emphasize that the R statistic introduced here sums over power in all modes, and the Monte Carlo simulations clearly demonstrate that this approach in fact does provide a very powerful tool for clustering studies.

As shown in Figure 2, the bulk of the previously allowed parameter space is now ruled out, and the new constraints are so tight that at 95% confidence, no more than $2\%/2 = 1\%$ of the burst sources can have repeated in the data set, assuming that repeaters have a brightness distribution typical of the entire catalog. Continued burst observations would allow further improvements in studies of this kind, but to really advance the field we must develop new missions that provide much better GRB localizations. While the forthcoming HETE mission (e.g., Ricker *et al.* 1992) may provide very accurate localizations, the event rate of HETE will probably be too low for recurrence to be detected. NASA has initiated several concept studies for new GRB missions, most recognizing the need for accurate positions. Future GRB observations may free us from the need to use statistical methods to answer such basic questions as burst recurrence, but for now we have no choice. The angular power method presented here should serve us well until the launch of a new generation of GRB experiments.

This work has been partially supported by European Union contract CHRX-CT93-0120, Deutsche Forschungsgemeinschaft grant SFB-375, NASA grant NAGW 5-1578, and NSF grant PHY94-07194 at the ITP.

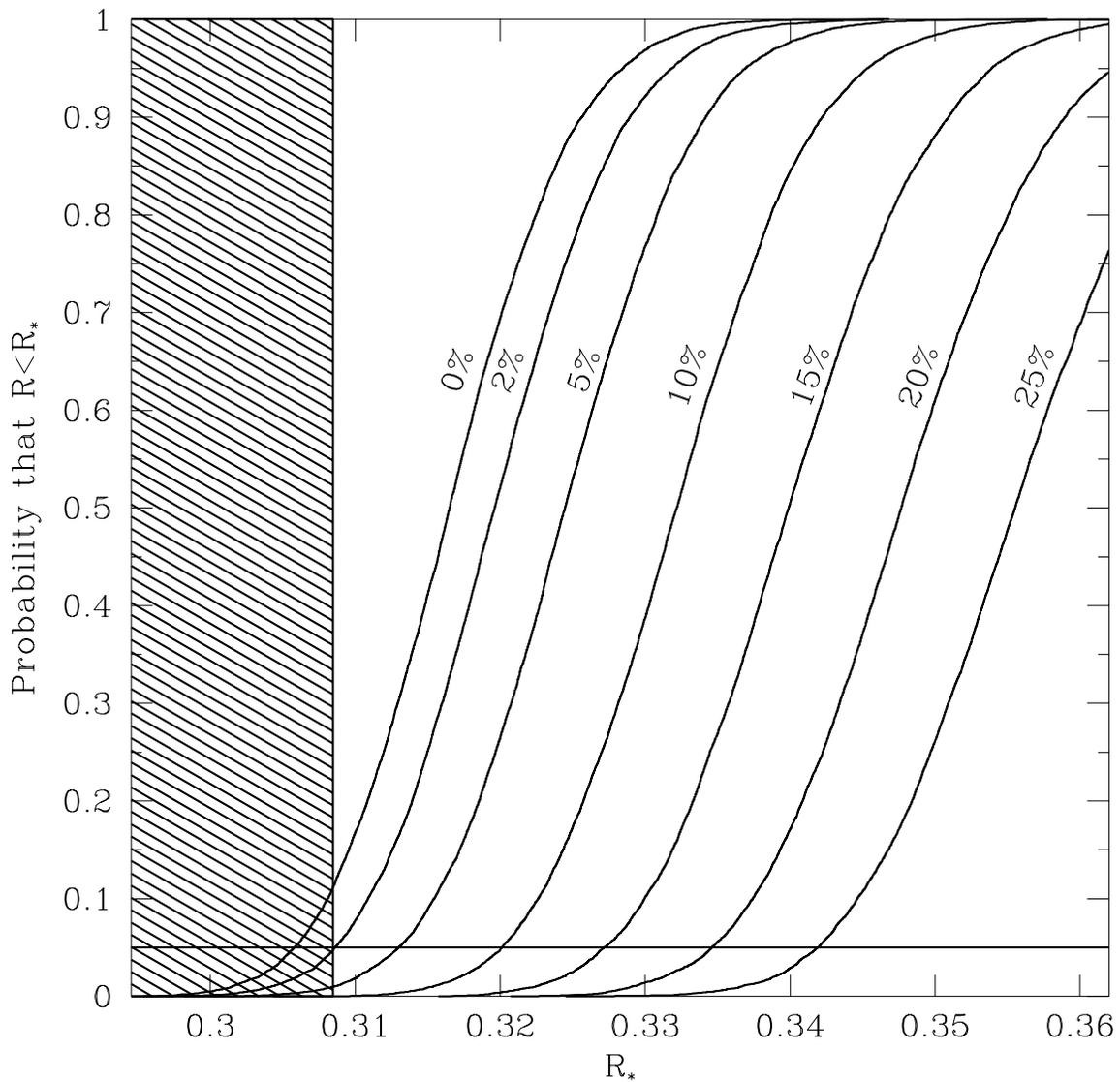

Figure 1: Monte Carlo results.
The cumulative probability distribution of our test statistic $R$ is shown, as determined from Monte Carlo simulations, for a range of repeater models. The models have a fraction $f$ of the bursts caused by repeating sources that are seen to burst twice ($\nu = 2$), where from left to right, $f = 0, 2\%, 5\%, 10\%, 15\%, 20\%$ and $25\%$. The shaded region contains the $R$-values smaller than that observed in the BATSE 3B data set, so the probability that a model is consistent with the data can be read off as the intersection of its curve with the vertical line. We see that $f = 2\%$ is ruled out at approximately $95\%$ confidence, since the intersection takes place near the horizontal $5\%$ line. (Note that since these curves depend on $N$ and $\sigma_k$, new Monte Carlo simulations must be made to analyze a different data set.)

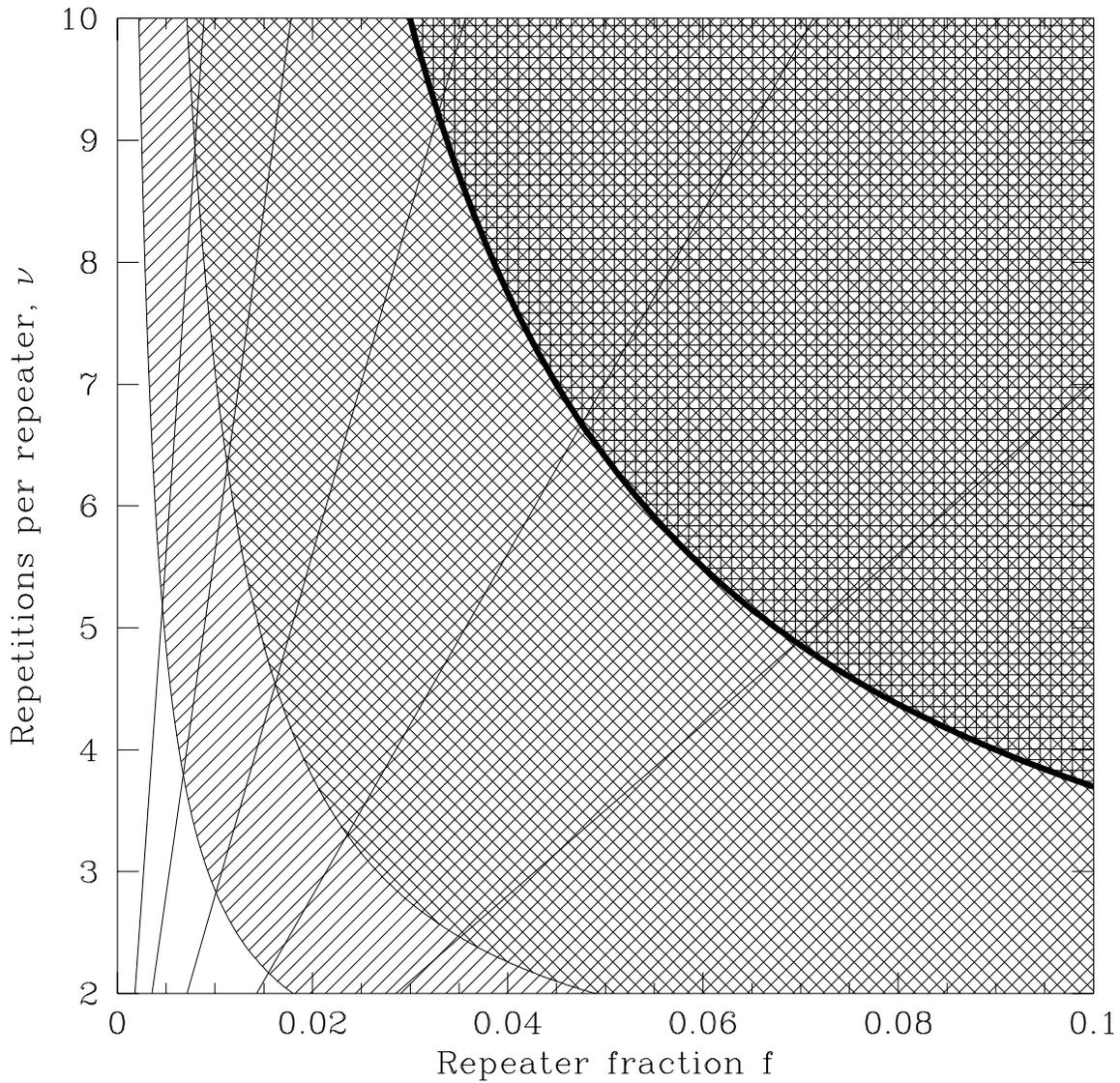

Figure 2: Excluded repeater models.
The shaded region of parameter space shows the repeater models that are ruled out by the BATSE 3B data at 95% confidence. The double-hatched region is excluded at 99% confidence. For comparison, the 99% confidence contour obtained using the correlation function (Meegan *et al.* 1995c) instead of our $R$-statistic is plotted (heavy solid line). From left to right, the straight lines correspond to models with 1, 2, 4, 8 and 16 repeating sources in the data set, respectively.